\documentclass{aa} 
\newcommand{\s}[1]{{_{\rm{#1}}}}

\usepackage{amsmath}
\usepackage{graphicx}
\usepackage{txfonts}
\usepackage{natbib}

\begin{document}
   \title{Numerical simulations of composite supernova remnants for small $\sigma$ pulsar wind nebulae}

   \subtitle{}

   \author{M.J. Vorster\inst{1}, S.E.S. Ferreira\inst{1}, O.C. de Jager\inst{1}, \and A. Djannati-Ata\"{i}\inst{2}
          }

   \offprints{M.J. Vorster}

   \institute{Centre for Space Research, School of
  Physics, North-West University, 2520 Potchefstroom, South Africa\\
       \email{12792322@puk.ac.za} \and CNRS, Universite Paris 7, Denis Diderot, F-75205, 75005, Paris, France}

   \date{Received xx/xx/xxxx}


  \abstract
   {Composite supernova remnants consist of a pulsar wind nebula located inside a shell-type remnant.  The presence of a shell has implications on the evolution of the nebula, although the converse is generally not true. 
}   
   {The purpose of this paper is two-fold.  The first aim is to determine the effect of the pulsar's initial luminosity and spin-down rate, the supernova ejecta mass,  and density of the interstellar medium on the evolution of a spherically-symmetric, composite supernova remnant expanding into a homogeneous medium.  The second aim is to investigate the evolution of the magnetic field in the pulsar wind nebula when the the composite remnant expands into a non-uniform interstellar medium.
}
   {The Euler conservation equations for inviscid flow, together with the magnetohydrodynamic induction law in the kinematic limit, are solved numerically for a number of scenarios where the ratio of magnetic to particle energy is $\sigma < 0.01$.  The simulations in the first part of the paper is solved in a one-dimensional configuration.  In the second part of the paper, the effect of an inhomogeneous medium on the evolution is studied using a two-dimensional, axis-symmetric configuration.
}
   {It is found that the initial spin-down luminosity and density of the interstellar medium has the largest influence on the evolution of the pulsar wind nebula.  The spin-down time-scale of the pulsar only becomes important when this value is smaller than the time needed for the reverse shock of the shell remnant to reach the outer boundary of the nebula.  For a remnant evolving in a non-uniform medium, the magnetic field along the boundary of the nebula will evolve to a value that is larger than the magnetic field in the interior.  If the inhomogeneity of the interstellar medium is enhanced, while the spin-down luminosity is decreased, it is further found that a magnetic "cloud" is formed in a region that is spatially separated from the position of the pulsar. 
}
{}
   \keywords{hydrodynamic --
                interstellar medium --
                supernova remnants                 
               }
\authorrunning{Vorster et al.}               
\titlerunning{Numerical simulations of composite SNRs}
  \maketitle
%

\section{Introduction}
It is well-known that a supernova explosion causes a shock wave that travels through the interstellar medium (ISM), creating a shell-like supernova remnant (SNR) that can be visible from radio to TeV energies \citep[e.g.][]{Woltjer-1972, H_Aharonian2006}.  This shock wave is not a single entity, but consists of three distinct parts: a forward shock that propagates into the ISM, a reverse shock, and a contact discontinuity separating the two shocks \citep[e.g.][]{Truelove-and-McKee-1999}.  Initially the reverse shock expands outwards along with the forward shock, but will eventually start to propagate into the interior of the SNR when a sufficient amount of ambient matter has been swept-up by the forward shock \citep{McKee1974}. 

The rate at which the SNR expands depends on a number of factors, including the mass of the stellar material ejected during the supernova explosion, $M\s{ej}$, the amount of kinetic energy transferred to the ejecta, typically of the order $E_{\rm{ej}} \sim 10^{51}$ erg, and the density of the interstellar medium $\rho_{\rm{ism}}$ \citep[e.g][]{Dwarkadas-and-Chevalier-1998,Truelove-and-McKee-1999,Tang-and-Wang-2005,Ferreira-and-de-Jager-2008}.  Additionally, the adiabatic index, $\gamma$, can also influence the evolution of a SNR due to the non-linear interaction of accelerated cosmic rays on SNR shocks \citep[e.g][]{Decourchelle-etal-2000,Ellison-etal-2004}.  For a comprehensive discussion of the evolution of SNRs, the papers of e.g. \citet{Sedov-1959,Woltjer-1972,Chevalier-1977}, and \citep{McKee-and-Truelove-1995} can be consulted.  

A possible by-product of a core-collapse supernova \citep[e.g.][]{Woosley-2005} is a pulsar, which in turn can lead to the presence of a second type of supernova remnant, commonly referred to as a pulsar wind nebula (PWN).  As the name suggests, the nebula is formed by a wind that originates from the pulsar and flows into the surrounding medium.  It is believed that this wind consists of relativistic leptons (electrons and positrons), and possibly a hadronic component \citep[e.g.][]{Horns-2006}, with the pulsar's magnetic field frozen into the plasma.  When the ram pressure of the wind is equal to the particle and magnetic pressure of the surrounding medium (created in earlier epochs of the pulsar wind), a termination shock will form \citep{Rees-and-Gunn-1974} where the particles are accelerated \citep[e.g.][]{Kennel_Coroniti}.  Downstream of the termination shock the accelerated leptons interact with the frozen-in magnetic field to produce synchrotron radiation ranging from radio to X-ray energies.  Additionally, the leptons can also inverse Compton scatter ambient photons to GeV and TeV gamma-ray energies.  The source of the ambient photons is typically the Cosmic Microwave Background Radiation, although infra-red radiation from dust, starlight, and even the synchrotron photons can also contribute.  The synchrotron and inverse Compton emission create a luminous nebula around the pulsar, with the pulsar continuously supplying energy to the PWN.  For a review of PWNe, the papers of \cite{Gaensler-2006}, \cite{Kargaltsev-2008} and \cite{de-Jager-2009} can be consulted.

An interesting subclass of SNRs are the systems consisting of a pulsar wind nebula located within a shell component.  Although the dynamics of the shell remnant is not influenced by the evolution of the PWN, the presence of a shell remnant can have a notable influence on the evolution of the PWN.  This one-sided interaction can be ascribed to the difference in energy content of the two components, with the shell remnant having a significantly larger amount of energy \citep[e.g.][]{Reynolds-and-Chevalier-1984}.  Simulations show that the evolution of a PWN can be divided into a number of stages \citep[e.g.][]{Reynolds-and-Chevalier-1984, Van-der-Swaluw-etal-2004}.  In the initial stage the PWN expands supersonically into the SNR ejecta until the reverse shock reaches the boundary of the PWN.  In the following stage the reverse shock compresses the PWN, with the interaction leading to the formation of a reflection wave that travels trough the PWN.  Calculations predict that the second stage should start at $\sim 10^3-10^4$ yr after the initial explosion \citep[e.g.][]{Reynolds-and-Chevalier-1984, Van-der-Swaluw-etal-2004}.  The final stage is characterised by the subsonic expansion of the PWN into the ejecta heated by the reverse shock.

Pulsars are often born with large kick-velocities, possibly resulting from an asymmetric supernova explosion.  The PWN is initially created at the centre of the SNR, but will be dragged along by the pulsar as it moves through the interior of the SNR \citep{Van-der-Swaluw-etal-2004}.  This reduces the distance between the shell and the PWN in the direction that the pulsar moves, thereby decreasing the time needed for the reverse shock to reach the forward boundary of the PWN.  For the opposite boundary, the reverse shock interaction time is correspondingly increased.  The different reverse shock interaction times leads the formation of a cigar-shaped nebula with the pulsar located at the edge  \citep{Van-der-Swaluw-etal-2004}.  The PWN continues to expand, while a bow-shock eventually forms at the nebula's edge closest to the pulsar when approximately two-thirds of the SNR radius has been traversed \citep{Van-der-Swaluw-etal-2003}.

Additionally, pulsars with a sufficiently large proper motion can break through the SNR shell and escape into the ISM, with the escape time for a $1000$ km $\rm{s}^{-1}$ pulsar estimated to be around $\sim 40000$ yr \citep{Van-der-Swaluw-etal-2003}.  A study of the velocity distribution of radio pulsars found characteristics at $90$ km $\rm{s}^{-1}$ and $500$ km $\rm{s}^{-1}$ \citep{Arzoumanian2002}.  It was further estimated that $15\%$ of the pulsars have velocities greater than $1000$ km $\rm{s}^{-1}$, and that only $10\%$ of pulsars younger than $20000$ yr will be found outside the SNR \citep{Arzoumanian2002}.  

Although the asymmetry observed in the morphology of PWNe can often be attributed to the large kick-velocities \citep[e.g.][]{Gaensler-2006}, interaction of the forward shock with a non-homogeneous ISM can lead to a similar morphology \citep{Blondin-etal-2001}.  It has been argued that the elongated morphology of (e.g.) Vela X and G09+0.1 is a result of the mentioned interaction, rather than a large proper motion of the pulsar \citep{Blondin-etal-2001}. 

In a previous paper, \cite{Ferreira-and-de-Jager-2008} calculated how the parameters $M_{\rm{ej}}$, $E_{\rm{ej}}$, $\rho_{\rm{ism}}$, and $\gamma$ influence the evolution of a SNR, with a part of their study focusing on the evolution of the shell in a non-uniform ISM.  In this work we extend their study by presenting hydrodynamic simulations of the evolution of a composite remnant containing a PWN for which the ratio of electromagnetic to particle energy is $\sigma < 0.01$.  One aim of the paper is to investigate how the pulsar's initial luminosity and spin-down time-scale, the mass of the supernova ejecta, and the density of the ISM influence the evolution of the PWN.  This parameter study is done for a spherically-symmetric remnant expanding into a homogeneous ISM.  The second aim of this paper is to illustrate the evolution of an axis-symmetric PWN in a non-homogeneous ISM, with an emphasis on the evolution of the magnetic field.  

The outline of the paper is as follows: Section 2 introduces the model that is used, with the spherically-symmetric, homogeneous ISM, results presented in Sections 3.  Section 4 motivates the use of the kinetic approach by illustrating that a more correct treatment of the magnetic field leads to similar results if $\sigma < 0.01$.  The axis-symmetric, inhomogeneous ISM results are presented in the Section 5, while the final section of the paper deals with the summary and conclusions.

\section{Model and parameters}
SNR evolution in uniform and non-uniform media have been modelled extensively by e.g. \cite {Tenorio-Tagle-etal-1991,Borkowski-etal-1997,Jun-and-Jones-1999}, while simulations focusing on selected aspects regarding the evolution of PWNe and composite remnants have been presented by e.g.\cite{Van-der-Swaluw-etal-2001,Blondin-etal-2001, Bucciantini-2002,Bucciantini-etal-2003, Van-der-Swaluw-2003,Del-Zanna-etal-2004,Bucciantini-etal-2005,Chevalier-2005}.  Similar to many of these studies, the present model is based on the well-known Euler equations for inviscid flow describing respectively the balance of mass, momentum and energy
\begin{equation}\label{hd}
\begin{split}
\frac{\partial}{\partial t}\rho+\nabla\cdot(\rho\textbf{v})&=0, \\
\frac{\partial}{\partial t}(\rho\textbf{v})+\nabla\cdot(\rho\textbf{v}\textbf{v}+P\textbf{I})&=0, \\
\frac{\partial}{\partial t}(\frac{\rho}{2}\textbf{v}^2+\frac{P}{\gamma-1})+\nabla\cdot(\frac{\rho}{2}\textbf{v}^2\textbf{v}+
\frac{\gamma\textbf{v}P}{\gamma-1})&=0.
\end{split}
\end{equation}
Here $\rho$ is the density, $\textbf{v}$ the velocity, and $P$ the pressure of the fluid.  The evolution of the magnetic field, $\textbf{B}$, is calculated by solving the induction equation
\begin{equation}\label{eq:induc}
\frac {\partial \textbf{B}}{\partial t} - \nabla \times (\textbf{v} \times \textbf{B}) =0,
\end{equation}
in the kinematic limit \citep{Scherer-and-Ferreira-2005a, Ferreira-and-Scherer-2006}.  In this approach, a one-sided interaction between the fluid and $\textbf{B}$ is assumed, with the effect of $\textbf{B}$ on the evolution of the flow neglected.  This approach is sufficient when the ratio of magnetic to particle energy is small, and will be further motivated in a later section. 

The Euler equations, (\ref{hd}), are solved numerically using a hyperbolic scheme based on a wave propagation approach, as discussed in detail in \cite{Leveque-2002} (see also \cite{Ferreira-and-de-Jager-2008} for additional details), while the scheme presented by \cite{Trac-and-Pen-2003} is used for the induction equation, (\ref{eq:induc}).  The model is solved in a spherical coordinate system with $0.01 \mbox{ pc} \le r \le 25$ pc and $0^{o} \le \theta \le 180^{o}$.  For PWNe evolving in a homogeneous medium (Section \ref{sec:EvoPWN}), the complexity of the problem reduces to a one-dimensional, spherically-symmetric scenario, and the corresponding results are discussed in this context.  For the simulations a $2400 \times 25$ grid is used for the homogeneous ISM, while the number of $\theta$ grid points is increased from $25$ to $85$ for the inhomogeneous ISM scenarios. 

In solving the equations, any relativistic effects are neglected, whereas a more realistic model should take into account that velocity of the particles between the pulsar and the PWN termination shock is highly relativistic.  However, since the post-shock flow must be subsonic, i.e. $\textbf{v}$ must be less than the sound speed in a relativistic plasma ($\sim c/\sqrt{3}$), smoothly decreasing towards the edge of the PWN, one expects that the fluid velocity in the PWN should decrease from mildly relativistic to non-relativistic.  

A second possible limitation arises from considering a one fluid scenario in which only a single value for the adiabatic index, $\gamma$, is specified.  A more correct model would treat the stellar ejecta and PWN as two separate fluids, with $\gamma\s{ej} = 5/3$ for the ejecta and $\gamma\s{pwn} = 4/3$ for the PWN.  \cite{Van-der-Swaluw-etal-2001} showed that this limitation can be compensated for in the one fluid model by setting $\gamma = 5/3$, while choosing the value for the mass deposition rate at the inner boundary of the computational domain, $\dot{M}_{\rm{pwn}}$, as 
\[
\sqrt{L(t)/\dot{M}_{\rm{pwn}}} = c,
\]
where $c$ is the speed of light.  Using a two-fluid, one-dimensional, relativistic magnetohydrodynamic (MHD) model, \cite{Bucciantini-etal-2003} found that the size of the PWN is only $\sim 10$\% smaller compared to one-fluid simulations using $\gamma = 5/3$.  

A further simplification in the present model is related to radiative losses being neglected in the evolution of the SNR.  The main emphasis of this study is on the interaction of the PWN with the reverse shock of the shell, and the subsequent evolution of the PWN.  The only limitation is that the time needed for the reverse shock to propagate back towards the PWN \citep[e.g.][]{Ferreira-and-de-Jager-2008}
\begin{multline}\label{eq:t_rs}
\hspace{-0.35cm}t_{\rm{rs}} = \\
4 \left(\frac{\rho_{\rm{ism}}}{10^{-24} \mbox{ g cm}^{-3}}\right)^{-0.33} \left(\frac{M_{\rm{ej}}}{3M_{\odot}}\right)^{0.75} \left(\frac{E_{\rm{ej}}}{10^{51} \mbox{ erg}}\right)^{-0.45} \left(\frac{\gamma}{5/3}\right)^{-1.5} \rm{ kyr} 
\end{multline}
must be smaller than the time when radiative losses become important  \citep{Blondin-etal-1998}
\begin{equation}\label{eq:t_pds}
t_{\rm{rad}} = 29 \left(\frac{E_{\rm{ej}}}{10^{51} \mbox{ erg}}\right)^{4/17}\left(\frac{\rho_{\rm{ism}}}{10^{-24} \mbox{ g cm}^{-3}}\right)^{-9/17} \rm{ kyr}.
\end{equation}
As discussed in a previous paragraph, the present model uses $\gamma=5/3$, together with the fiducial value $E_{\rm{ej}}=10^{51}$ $\rm{erg}$.  The time-scales (\ref{eq:t_rs}) and (\ref{eq:t_pds}) are thus reduced to a function of the ejecta mass and ISM density.  For any permutation of the values $M_{\rm{ej}}$ and $\rho_{\rm{ism}}$ chosen from the ranges $M\s{ej}=4M_{\odot}-16M_{\odot}$ and $\rho_{\rm{ism}}=10^{-25}-10^{-23} \mbox{ g cm}^{-3}$, one has $t_{\rm{rs}} < t_{\rm{rad}}$. 

One shortcoming of the kinematic approach is that it does not reproduce the required PWN morphology in the initial evolutionary phase.  Axis-symmetric, two-dimensional MHD simulations, extending up to $t=1000$ yr, have shown that the magnetic field in the PWN leads to an elongation of the nebula from a spherical to an elliptic-like shape, with a ratio of $1.3-1.5$ between the semi-major and semi-minor axes \citep{Van-der-Swaluw-2003}.  This non-spherical structure will induce time delays between the interaction of the reverse shock with the polar/equatorial regions of the PWN.  A similar feature is predicted by the axis-symmetric, two-dimensional, relativistic MHD simulations of \cite{Del-Zanna-etal-2004}. However, the latter authors found that the elongation increases to a maximum value at around $t=400$ yr, before decreasing again.  When the ratio of magnetic to particle energy is $\sigma = 0.01$, the elongation ratio decreases to a value that is less than 1.2 after $t=1000$ yr.  Since the simulations only extend up to $t=1000$ yr, it may be possible that the elongation will reduce further.  Additionally, \cite{Del-Zanna-etal-2004} also found that the elongation ratio reaches unity after $t=1000$ yr for the case $\sigma=0.003$.  Note that the simulations presented in this paper are specifically for scenarios where $\sigma < 0.01$.     

It is unclear what role the asymmetric structure plays in the long-term evolution of the PWN since this has, to the authors' knowledge, not been studied.  It is heuristically argued that the effect would not be as pronounced as one would expect.  When the reverse shock reaches the polar boundary, the nebula will be compressed in these regions and the thermal pressure increased.  Due to the large sound speed in a PWN, the nebula will reach pressure equilibrium in a very short time.  The larger pressure in the PWN will lead to an accelerated expansion in the regions where the reverse shock has not yet interacted with the PWN, specifically in the equatorial regions.  This, in turn, reduces the time needed for the reverse shock to reach the equator of the PWN, leading to a more "symmetric" interaction between the PWN and reverse shock.  Alternatively, it may also be argued that the evolution of the PWN is significantly more dependent on factors such as a large pulsar kick velocity or and inhomogeneous ISM.

Simulations have shown that the interaction of the reverse shock with the PWN leads to the formation of Raleigh-Taylor instabilities \citep[e.g.][]{Blondin-etal-2001, Van-der-Swaluw-etal-2004, Bucciantini-etal-2004a}.  Additionally, the same instabilities are expected to occur at the forward shock of young SNRs \citep[e.g.][]{Blondin-etal-2001}.  An initial test of the present numerical scheme found that these instabilities can be produced if a sufficiently fine computational grid is used.  However, to avoid numerical dissipation, the grid has to be refined to such a degree that the code becomes extremely computationally intensive.  Since the aim of the present paper is to simulate the large-scale evolution and structure of the PWN, any further investigation and discussion of these instabilities are neglected.

In the initial configuration of the system \citep[e.g.][]{Blondin-and-Ellison-2001,Van-der-Swaluw-etal-2001,Bucciantini-etal-2003, Del-Zanna-etal-2004}
the SNR is spherical with a radius $R_{\rm{ej}}=0.1$ pc and a radially increasing velocity profile
\begin{equation}
v=\frac{r}{t}= \frac{v\s{ej}r}{R\s{ej}}, 
\end{equation}
with
\begin{equation}\label{speed}
v_{\rm{ej}}=\sqrt{\frac{10}{3}\frac{E_{\rm{ej}}}{M_{\rm{ej}}}},
\end{equation}
while the density of the ejecta is uniform in the interior of the SNR 
\begin{equation}\label{density}
\rho_{\rm{ej}}=\frac{3M_{\rm{ej}}}{4 \pi R_{\rm{ej}}^3}. 
\end{equation}
Note that in the initial condition the PWN is absent, with the particles injected by the pulsar included after the first time-step.

For the boundary conditions \citep[e.g.][]{Blondin-and-Ellison-2001,Van-der-Swaluw-etal-2001,Bucciantini-etal-2003, Del-Zanna-etal-2004}, the speed of the particles injected at the inner boundary of the computational domain is equal to the speed of light, $c$, while the density is calculated from the mass deposition rate 
\citep{Van-der-Swaluw-2003}
\begin{equation}
 \dot{M}_{\rm{pwn}} = \frac{2L}{c^2},
\end{equation}
where
   \begin{equation}\label{L_t}
     L(t) = L_0\left( 1+\frac{t}{\tau}\right)^{-\left(n+1\right)/\left(n-1\right)},
   \end{equation}
is the spin-down luminosity of the pulsar.  The variable $\tau$ is known as the spin-down time scale, and is defined by 
   \begin{equation}\label{eq_tau}
     \tau = \frac{4\pi^2 I}{\left(n-1\right)P_0^2 L_0},
   \end{equation}  
  with $L_0$ and $P_0$ respectively the initial luminosity and spin period, $I$ the moment of inertia, and $n$ the pulsar braking index.  
  
The magnetic field at the inner boundary is assumed to be purely azimuthal, in accordance with theory \citep[e.g.][]{Rees-and-Gunn-1974} and previous PWN models \citep[e.g.][]{Van-der-Swaluw-2003}, while an expression for the time evolution of the magnetic field strength, $B(t)$, at the inner boundary is derived as follows: the spin-down luminosity, $L(t)$, at any given position, $r$, upstream of the pulsar wind termination shock can be written as \citep{Kennel_Coroniti}  
  \begin{equation}\label{L_upstream}
    L(t) = c r^2 B(t)^2 \frac{1+\sigma}{\sigma},
  \end{equation}
where $\sigma$ is the ratio of electromagnetic to particle energy.  Equation (\ref{L_upstream}) is effectively a statement of the conservation of energy, and is valid for any value of $\sigma$.  Since Equation (\ref{L_upstream}) should hold for all times, $t$, one may also state that  
  \begin{equation}\label{L_upstream_0}
    L_0 = c r^2 B_0^2 \frac{1+\sigma}{\sigma}.
  \end{equation}
Inserting (\ref{L_upstream_0}) into (\ref{L_t}) leads to the expression 
    \begin{equation}\label{L_B_0}
      L(t) = c R^2 B_0^2 \frac{1+\sigma}{\sigma}\left( 1+\frac{t}{\tau}\right)^{-\left(n+1\right)/\left(n-1\right)}.
    \end{equation}
  To calculate the time-dependence of $B$, the right-hand sides of Equations (\ref{L_upstream}) and (\ref{L_B_0}) are equated, while it is assumed that $\sigma$ has no time dependence at the inner boundary, leading to the result
    \begin{equation}\label{B_t}
      B(t) = B_0\left( 1+\frac{t}{\tau}\right)^{-\left(n+1\right)/\left(n-1\right)}.
    \end{equation}    
In addition, the initial value of the magnetic field at the inner boundary of the computational domain is normalised to $B_0=1$.  This is possible since the magnetic field (in the present) model does not influence the dynamics of the PWN. 
  
For the simulations, the initial luminosity ranges between $L_0=10^{37}-10^{40}$ erg $\rm{s}^{-1}$, and the spin-down time-scale between $\tau=500-5000$ yr.  Based on observations, (e.g.) \cite{Zhang2008, Fang2010, Tanaka2011} have derived similar $L_0$ and $\tau$ values for a number of PWNe.  Core-collapse supernovae (where pulsars are produced as a by-product) are expected to take place for stars with masses $\gtrsim 8M_{\odot}$ \citep[e.g.][]{Woosley-2005}.  As an example, a value of $M_{\rm{ej}}\sim 11M_{\odot}-15M_{\odot}$ has been derived for SN 1987A \citep[e.g.][]{Shigeyama-1990}.  In the simulations the ejecta mass ranges between $M_{\rm{ej}} = 8M_{\odot}-12M_{\odot}$.  Furthermore, core-collapse supernovas occur in star forming regions \citep[e.g.][]{Woosley-2005} in the ISM with densities ranging from $\rho_{\rm{ism}}=10^{-26}-10^{-23} \mbox{ g cm}^{-3}$ \citep[e.g.][]{Thornton-etal-1998}, and a similar range chosen for the model. Lastly, the fiducial values $E\s{ej}=10^{51}$ erg and $n=3$ are used for the simulations.

\cite{Van-der-Swaluw-and-Wu-2001} investigated the role of $P_0$ on the evolution of a composite remnant, and concluded that if $P_{0} \leq 4$ ms, the energy in the pulsar wind exceeds the total mechanical energy of the SNR, resulting in the SNR being blown away.  This result is indirectly supported by the findings of \cite{Kaspi_birth}.  Using alternative means, these authors calculated $P_0$ for a number of galactic pulsars, with $P_0 = 11$ ms being the lowest value found.  The values of $L_0$ and $\tau$, related through (\ref{eq_tau}), must therefore be chosen in such a way as to not result in an unrealistic $P_0$, with the definition of "realistic" being (loosely) defined as $P_0 \gtrsim 4$ ms.  Using the fiducial values $I = 10^{45} \mbox{ g cm}^2$ and $n = 3$ for the moment of inertia and braking index of the pulsar, the smallest $P_0$ value for the present simulations is found for the $L_0=10^{40}$ erg, $\tau=5000$ yr scenario, where $P_0=3.5$ ms.  Although this scenario is technically below the accepted limit, it is nevertheless included for the purpose of illustration.

\section{Evolution of a PWN inside a Spherically-symmetric SNR}\label{sec:EvoPWN}

In this section the effect of the parameters $L_0$, $\tau$, $M\s{ej}$, and $\rho\s{ism}$ on the evolution of a spherically-symmetric PWN in a homogeneous ISM is investigated.  In order to better compare the various scenarios, it is useful to define a basis set of parameters, chosen as $L_0=10^{39}$ erg $\rm{s}^{-1}$ for the initial luminosity, $\tau=3000$ yr for the spin-down time-scale, $M_{\rm{ej}} = 12M_{\odot}$ for the mass of the stellar ejecta, and $\rho_{\rm{ism}}=10^{-25}$ $\mbox{ g cm}^{-3}$ as the density of the ISM.

\subsection{The outer boundary of the PWN}

\begin{figure*}[!t]
   \centering
  \includegraphics[width=25pc]{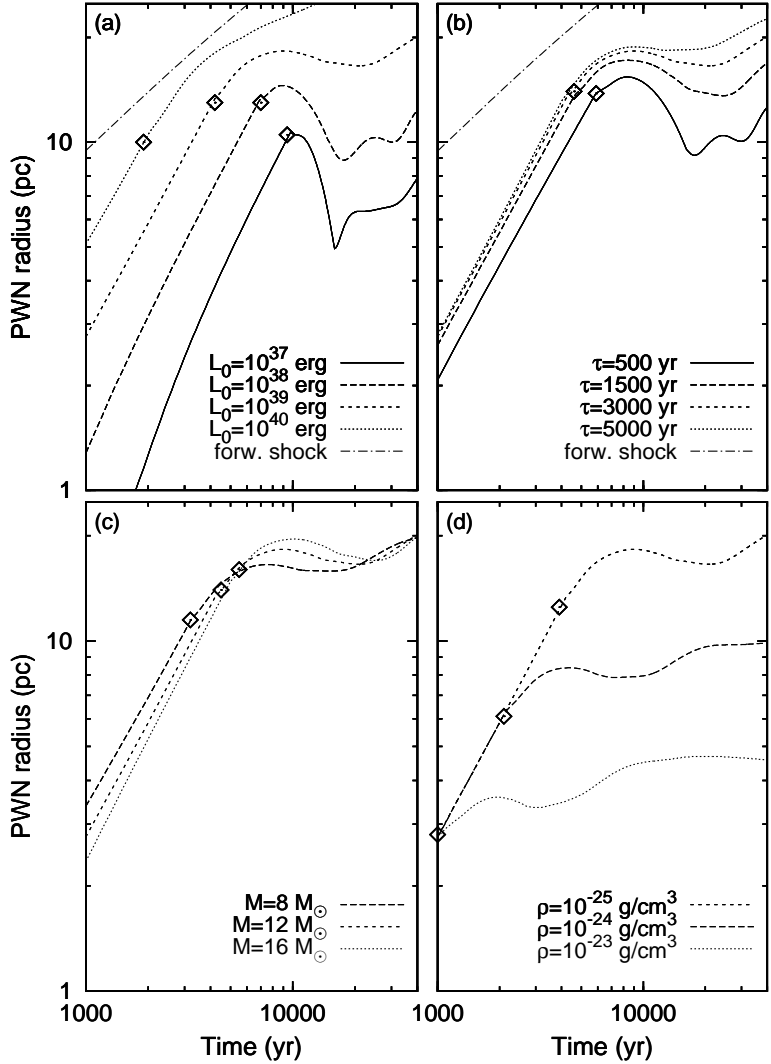}
   \vspace{-0.4cm}
   \caption{Model solutions for the evolution of the outer boundary of a spherically-symmetric PWN in a homogeneous ISM.  A basis parameter set is chosen as $L_0=10^{39}$ erg $\rm{s}^{-1}$, $\tau=3000$ yr, $M_{\rm{ej}} = 12M_{\odot}$, and $\rho_{\rm{ism}}=10^{-25}$ $\mbox{ g cm}^{-3}$, with the solution represented by a short-dashed line in Figures (a)-(d).  The other solutions were obtained by using the basis parameters, while varying the parameters listed in the legends of the figures.  The $\Diamond$ symbol represents the time when the reverse shock reaches $R\s{pwn}$.  Also shown in Figures (a) and (b) is the evolution of the forward shock of the SNR.}
              \label{fig:pwn_radius}%
    \end{figure*}

Figure shows the effect of the parameters on the evolution of the outer boundary of the nebula, $R\s{pwn}$.  The solutions were obtained with the basis values, only varying the appropriate variable as stated in the legends of the figures.  The basis solution in the various panels of Figure \ref{fig:pwn_radius} is represented with a short-dashed line.  Figures \ref{fig:pwn_radius}(a)-(b) also shows the evolution of the forward shock of the SNR.  Since the forward shock is not dependent on the values of $L_0$ and $\tau$, the evolution is the same for all scenarios shown in these two panels.  For Figures \ref{fig:pwn_radius}(c)-(d), the evolution of the forward shock is the same as that described in \cite{Ferreira-and-de-Jager-2008}.  Before the interaction with the reverse shock (indicated by the $\Diamond$ symbols), the rate of expansion of the PWN can be described by a power-law function, $R\s{pwn} \propto t^{1.1-1.3}$, similar to the prediction $R\s{pwn} \propto t^{1.2}$ made by \cite{Reynolds-and-Chevalier-1984}.   

For the different $L_0$ scenarios in Figure \ref{fig:pwn_radius}(a), a larger PWN is predicted when $L_0$ is increased.  This difference in PWN size can be attributed to the larger $L_0$ scenarios having a faster expansion rate at the earliest times, $t < 100$ yr, with the expansion rate eventually evolving to $R\s{pwn} \propto t^{1.2}$.  Larger $L_0$ values not only leads to a larger PWN, but also imply that more particles are injected into the PWN per time interval, thereby increasing the pressure in the PWN.  For the $L_0=10^{40}$ $\mbox{erg s}^{-1}$ scenario, the interaction with the reverse shock does not lead to a compression of the PWN, but only a slower (subsonic) expansion rate.  For the $L_0=10^{37}$ $\mbox{erg s}^{-1}$ scenario (where the PWN pressure is smaller), the reverse shock noticeably compresses the PWN, with the post-interaction expansion rate having an oscillatory nature.

At times $t \lesssim \tau$, the value of $L_0$ is almost constant, and any variation in $\tau$ should not affect the evolution of $R\s{pwn}$.  This can be seen from the results in Figure \ref{fig:pwn_radius}(b) where the evolution of $R\s{pwn}$ at times $t \lesssim 5000$ yr is almost identical for the various scenarios.  The exception is the $\tau=500$ yr scenario where a smaller PWN is predicted.  The largest influence of $\tau$ is seen after the interaction with the reverse shock.  Since a smaller $\tau$ leads to a faster decrease in $L(t)$, the PWN pressure should correspondingly decrease, leading to a larger compression of the PWN.  However, as shown in Figure \ref{fig:pwn_radius}(b), this effect is not as prominent for a PWN where the reverse shock time-scale is of the order of $\tau$.

A larger mass for the stellar ejecta, $M\s{ej}$, leads to an increase in the density of the matter that the PWN expands into, and one would expect a slower expansion rate for $R\s{pwn}$.  However, as shown in Figure \ref{fig:pwn_radius}(c), the expansion rate is almost identical for the various $M\s{ej}$ scenarios.  Similar to Figure \ref{fig:pwn_radius}(a), a larger PWN size for smaller $M\s{ej}$ values is a result of a faster expansion in the earliest years of the PWN.  Although the reverse shock time-scale is reduced for a larger PWN, this effect is partially cancelled since a larger $M\s{ej}$ value also increases the same time-scale.

Again using (\ref{eq:t_rs}), it can be seen that a larger value of $\rho\s{ism}$ markedly reduces the time needed for the reverse shock to reach the PWN.  This decreases the duration of the supersonic expansion phase of $R\s{pwn}$, leading to a PWN that is noticeably smaller, as shown in Figure \ref{fig:pwn_radius}(d).  From the $\rho_{\rm{ism}}=10^{-24}$ $\mbox{ g cm}^{-3}$ and $\rho_{\rm{ism}}=10^{-23}$ $\mbox{ g cm}^{-3}$ scenarios it can also be seen that the PWN eventually reaches a stage where $R\s{pwn}$ approaches a constant value.

\subsection{The evolution of the termination shock radius}

\begin{figure*}[t]
   \centering
  \includegraphics[width=25pc]{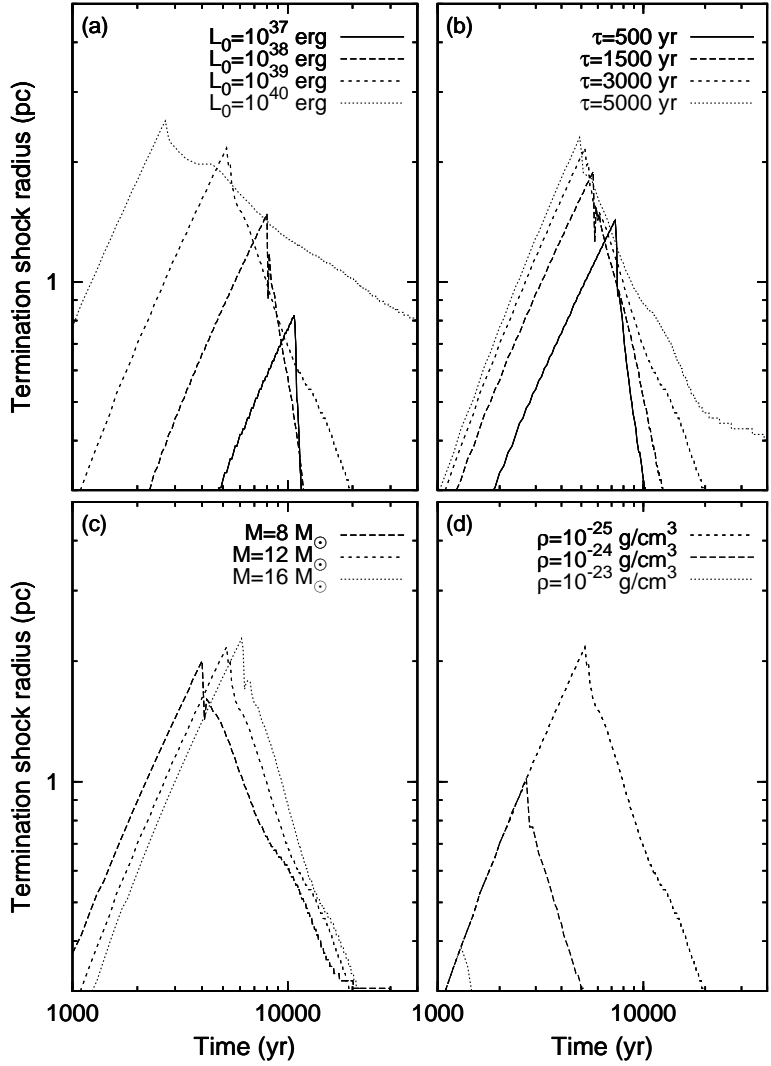}
   \vspace{-0.4cm}
   \caption{The same as Figure \ref{fig:pwn_radius}, but for the termination shock radius, $R\s{ts}$.}
              \label{fig:ts_radius}%
    \end{figure*}

The next quantity investigated is the termination shock radius, $R\s{ts}$.  Figure \ref{fig:ts_radius} is the corresponding evolution of $R\s{ts}$ for the scenarios shown in Figure \ref{fig:pwn_radius}.  Note that the striations shown in Figure \ref{fig:ts_radius} is of a numerical origin, resulting from the chosen grid resolution. For the various scenarios, $R\s{ts}$ expands until the PWN encounters the reverse shock.  The subsequent slower (subsonic) expansion and compression leads to an increase in the pressure in the PWN, resulting in $R\s{ts}$ being pushed back towards the pulsar.  The effect of the various parameters on the evolution of $R\s{ts}$ mirrors the results of Figure \ref{fig:pwn_radius} as a larger PWN also has a larger $R\s{ts}$ before the reverse shock interaction.  After the reverse shock interaction, a faster decrease in $R\s{ts}$ is predicted for a larger compression of the PWN.

\subsection{The evolution of the average magnetic field}

\begin{figure*}[t]
   \centering
  \includegraphics[width=25pc]{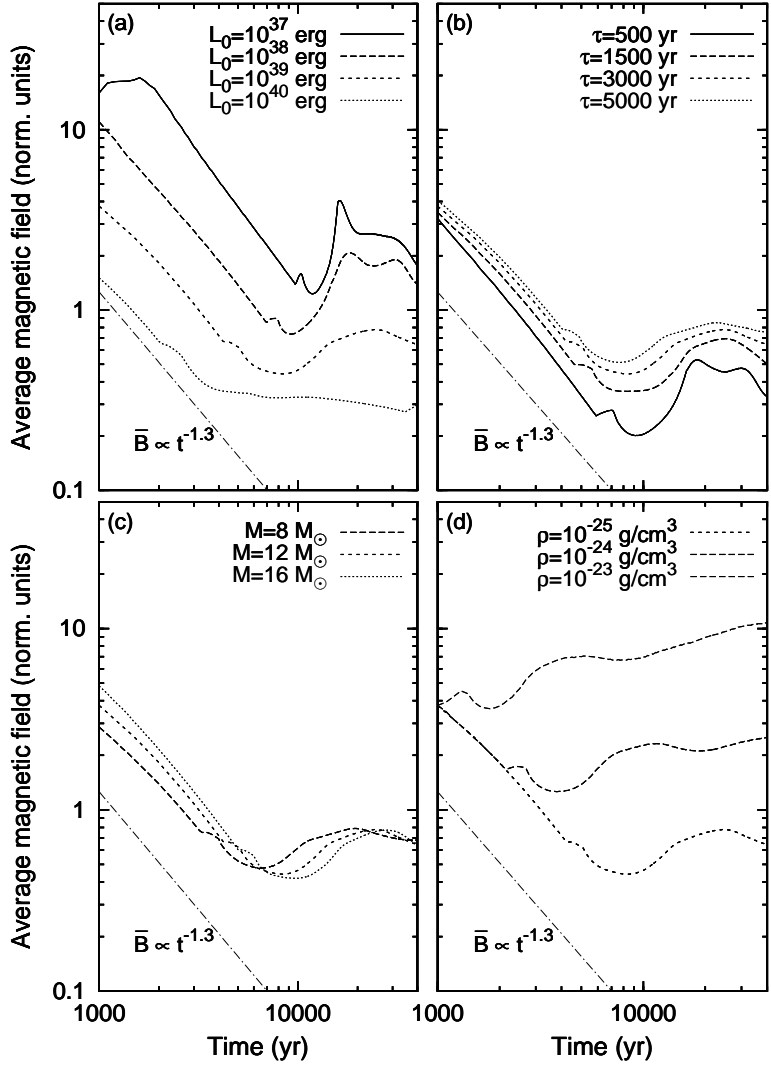}
   \vspace{-0.4cm}
   \caption{The same as Figure \ref{fig:pwn_radius}, but for the average magnetic field in the PWN.  Note that the magnetic field strength at the inner boundary of the computational domain has been normalised to unity.}
              \label{fig:b_field}%
    \end{figure*}

Figure \ref{fig:b_field} shows the temporal evolution of the average magnetic field, $\bar{B}$, in the PWN, defined as the region $R\s{ts} \le r \le R\s{pwn}$.  The solutions correspond to those shown in Figures \ref{fig:pwn_radius} and \ref{fig:ts_radius}.  If the magnetic energy in the PWN is not transferred to the particles, then the conservation of magnetic flux implies that the evolution of $\bar{B}$ should be related to the evolution of $R\s{pwn}$.  A comparison of Figures \ref{fig:pwn_radius} and \ref{fig:b_field} shows that an increase in $R\s{pwn}$ leads to a decrease in $\bar{B}$, with the opposite also being true.  Furthermore, the rate of change in $R\s{pwn}$ is also reflected in the rate of change in $\bar{B}$.

Before the interaction with the reverse shock, it is possible to derive a simple time-dependence for $\bar{B}$ since the expansion of $R\s{pwn}$ is described by a power-law.  
From the conservation of magnetic flux it follows that 
\begin{equation}\label{eq:B_tot}
\int_0^t \eta Ldt' = V\s{pwn}\frac{\bar{B}^2}{8\pi},
\end{equation}
where $\eta$ is the fraction of spin-down luminosity converted to magnetic energy and $V\s{pwn}\propto R\s{pwn}^3$ is the volume of the PWN.  Note that the integral in (\ref{eq:B_tot}) represents the total amount of magnetic energy injected into the PWN over the time interval $0 \le t' \le t$.  If $t \lesssim \tau$, then $L$ can be approximated as constant, reducing the integral in (\ref{eq:B_tot}) to $L_0t$.  Lastly, using $R\s{pwn}\propto t^{\alpha}$ it becomes possible to reduce (\ref{eq:B_tot}) to
\begin{equation}
\bar{B} \propto t^{(3\alpha-1)/2}.
\end{equation}
For the values $\alpha=1.1-1.3$, as derived from Figure \ref{fig:pwn_radius}, the evolution of the average magnetic field in Figure \ref{fig:b_field} (before the interaction with the reverse shock) can be described by $\bar{B} \propto t^{1.1-1.5}$.  The same arguments have also been used by \cite{Reynolds-and-Chevalier-1984} to derive the evolution of $\bar{B}$.  For reference, the line $\bar{B} \propto t^{-1.3}$ is also shown in Figure \ref{fig:b_field}.  For subsequent times, it is difficult to characterise the evolution of $\bar{B}$ with a simple expression as the evolution of $R\s{pwn}$ becomes more complex.

\subsection{Radial profiles of the fluid quantities}

\begin{figure}[!t]
   \centering
   \vspace{13.1cm}
  \includegraphics{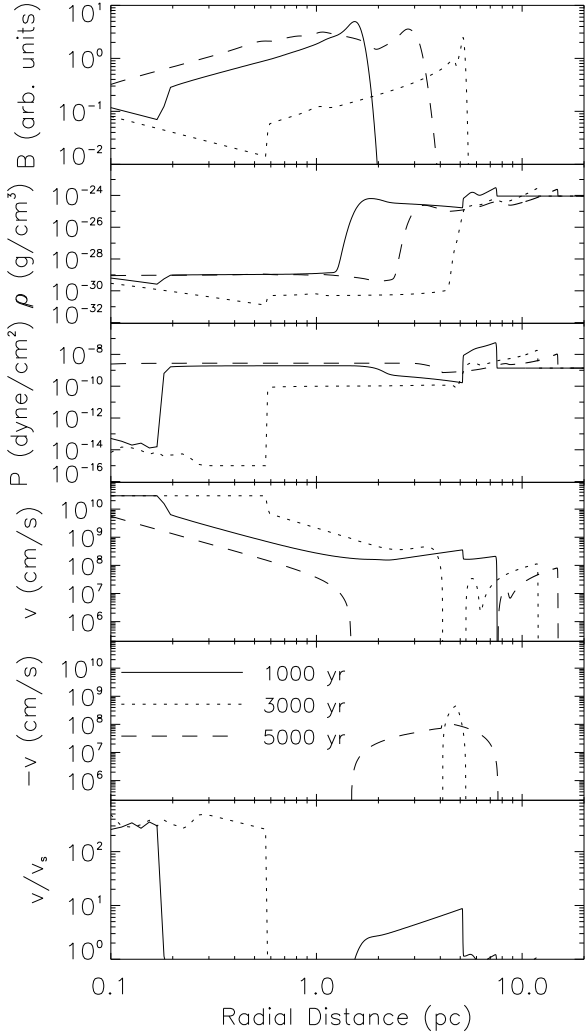}
   \caption{From top to bottom: radial profiles of the magnetic field strength $B$, density $\rho$, pressure $P$, speed $v$ ($-v$ indicates a velocity in the opposite direction), and Mach number at three different times: 1000 yr (solid line), 3000 yr (dotted line), and 5000 yr (dashed line).  The profiles shown are for the $L_{0}=10^{38}$ $\mbox{erg s}^{-1}$, $\tau$ = 3000 yr and $\rho_{\rm{ism}}=10^{-24}$ $\mbox{g cm}^{-3}$ scenario.  Note that $B$ has again been normalised to unity at the inner boundary.}
              \label{p4versameling2}%
    \end{figure}

It is also useful to investigate the evolution of the different fluid quantities, especially for times before and after the interaction of $R_{\rm{pwn}}$ with the reverse shock. The radial profiles of the fluid quantities at three different times are shown in Figure \ref{p4versameling2}, corresponding to the scenario $L_{0}=10^{38}$ $\mbox{erg s}^{-1}$, $\tau$ = 3000 yr and $\rho_{\rm{ism}}=10^{-24}$ $\mbox{g cm}^{-3}$.

The magnetic profile (top panel) is useful for identifying the various components of the PWN.  The radial distance where $B$ falls off to zero indicates the position of $R_{\rm{pwn}}$, while a sharp increase in $B$ indicates the position of $R_{\rm{ts}}$ (a shock compresses $B$ over a very small region).  The $B\propto 1/r$ decrease before the termination shock is due to the assumption of an azimuthal magnetic field at the inner boundary of the computational domain.  An interesting feature of the magnetic field is the increase in strength after $R\s{ts}$, related to the decrease in velocity (see Panel 4).  A similar effect occurs in our local heliosphere, with the enhancement known as the Axford-Cranfill effect \citep[][]{Cranfill-1971, Axford-1972}. A magnetic wall emerges from the solar wind termination shock, arising from the amplification of the heliospheric magnetic field's azimuthal component, which in turn is caused by the flow deceleration and the convection to higher latitudes \citep[e.g.][]{Zank-1999}.  

Using the density profile, Panel 2, allows one to distinguish the different components of the composite remnant.  The low density region represents the PWN, and the larger density region the ejecta.  The $t=1000$ yr profile shows two peaks at $r \sim 5$ pc and $r\sim 7$ pc, with these peak respectively representing the reverse and forward shock of the SNR.  For larger $r$ values, the constant density region delineates the ISM.  Comparing the density profiles of the various times with the position of $R_{\rm{pwn}}$, as defined by Panel 1, shows that there is a region just behind $R_{\rm{pwn}}$ where the density increases significantly.  This can be heuristically viewed as being a result of $R_{\rm{pwn}}$ interacting with the denser ejecta, leading to a pile-up of PWN material.  Such an explicit distinction is, however, difficult to make in a one-fluid model.   

Panel 3 shows that the pressure behind the reverse shock is lower than both the pressure in the PWN and the SNR shock.  It is the pressure difference between the ejecta and shock that eventually leads to the reverse shock propagating back towards the pulsar.  After the interaction with the reverse shock, the difference in pressure between the PWN and the ejecta decreases, reducing the PWN expansion rate (see Figure \ref{fig:pwn_radius})

Apart from compressing the PWN (and consequently enhancing $B$), the interaction between the reverse shock and $R_{\rm{pwn}}$ drives a reflection wave through the PWN, as illustrated in Panel 5.  Lastly, Panel 6 shows that the outer edge of the PWN initially expands supersonically into the ejecta, becoming sub-sonic once the reverse shock has passed.

\section{Motivation for the Kinematic Appraoch}\label{sec:MHD}

A more correct treatment of the present problem should take into account the effect that the magnetic pressure has on the evolution of the PWN.  An ideal approach would be to use an axis-symmetric, 2.5-dimensional MHD model, similar to the one presented by \cite{Van-der-Swaluw-2003}.  However, since these models tend to be computationally expensive, it becomes difficult to do an extensive parameter as presented in Figures \ref{fig:pwn_radius}-\ref{fig:b_field}.  The aim of this section is to motivate that a hydrodynamic model with a kinematic approach is a useful approximation when the ratio of magnetic to particle energy is below $\sigma < 0.01$.  To obtain an estimate of the accuracy, the one-dimensional model presented in the previous section is used, with the difference that the magnetic pressure is calculated at every time step, and added to the total pressure.  Similar to the kinematic scenario, this approach has the limitation that it does not produce the required deviations from spherical symmetry. Even though the same model is used, including the effect of the magnetic pressure significantly increases the computational time.  The simulations are therefore limited to the single scenario $L_{0}=10^{40}$ $\mbox{erg s}^{-1}$, $\tau$ = 300 yr, $M_{\rm{ej}}=8M_{\odot}$ and $\rho_{\rm{ism}}=10^{-24}$ $\mbox{g cm}^{-3}$, with the PWN evolution only calculated up to $t=1000$ yr.

\begin{figure*}[t]
\begin{minipage}[b]{.48\linewidth}
   \centering
  \includegraphics[width=20pc]{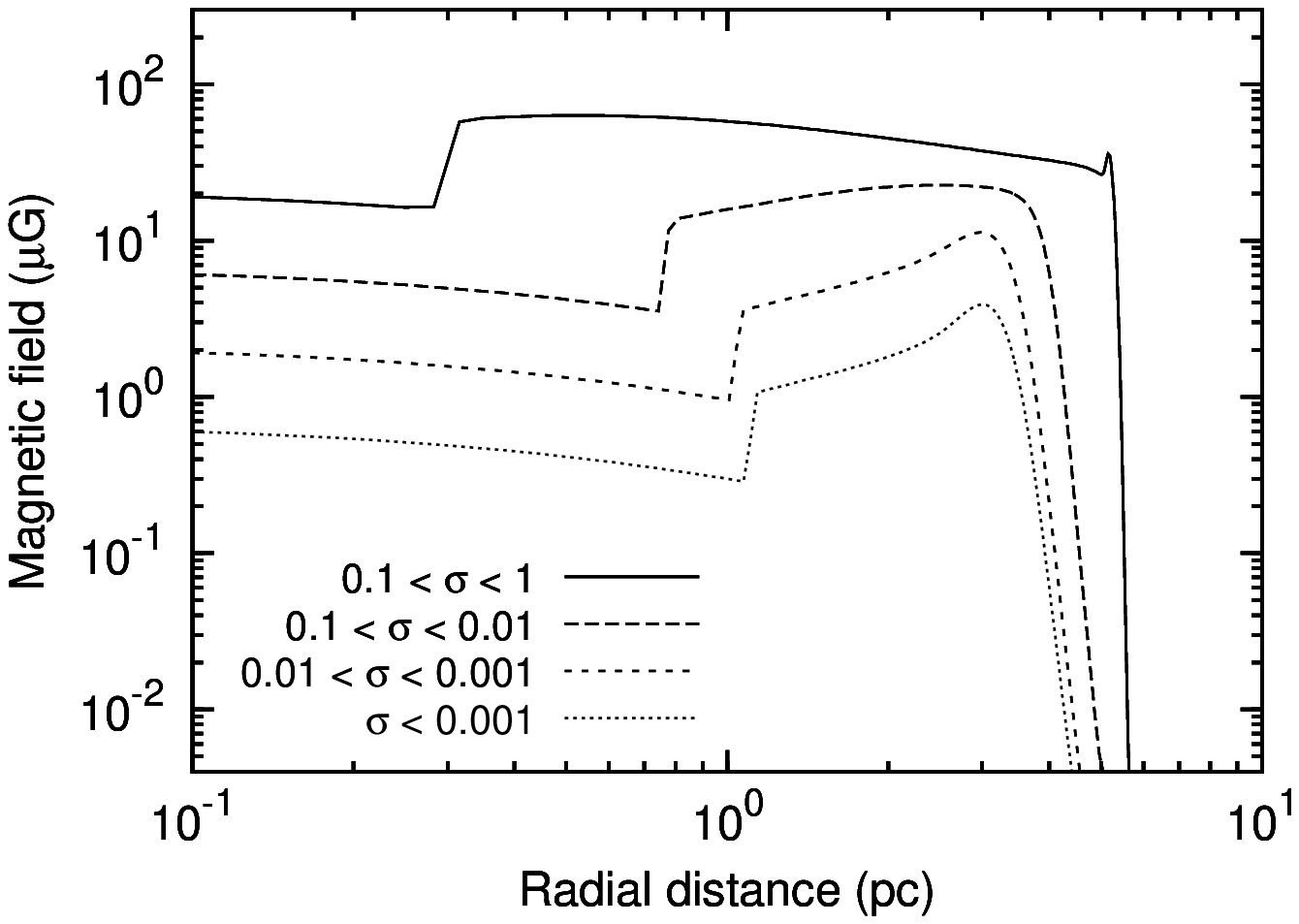}
   \caption{Radial profile of the PWN magnetic field after $t=1000$ yr for various values of $\sigma$.  A range is specified as the values are not constant in the PWN.  The plots are for the $L_{0}=10^{40}$ $\mbox{erg s}^{-1}$, $\tau$ = 300 yr, $M_{\rm{ej}}=8M_{\odot}$ and $\rho_{\rm{ism}}=10^{-24}$ $\mbox{g cm}^{-3}$ scenario. }
              \label{bpres}%
\end{minipage}%
\hspace{0.5cm}
\begin{minipage}[b]{.48\linewidth}
   \centering
  \includegraphics[width=20pc]{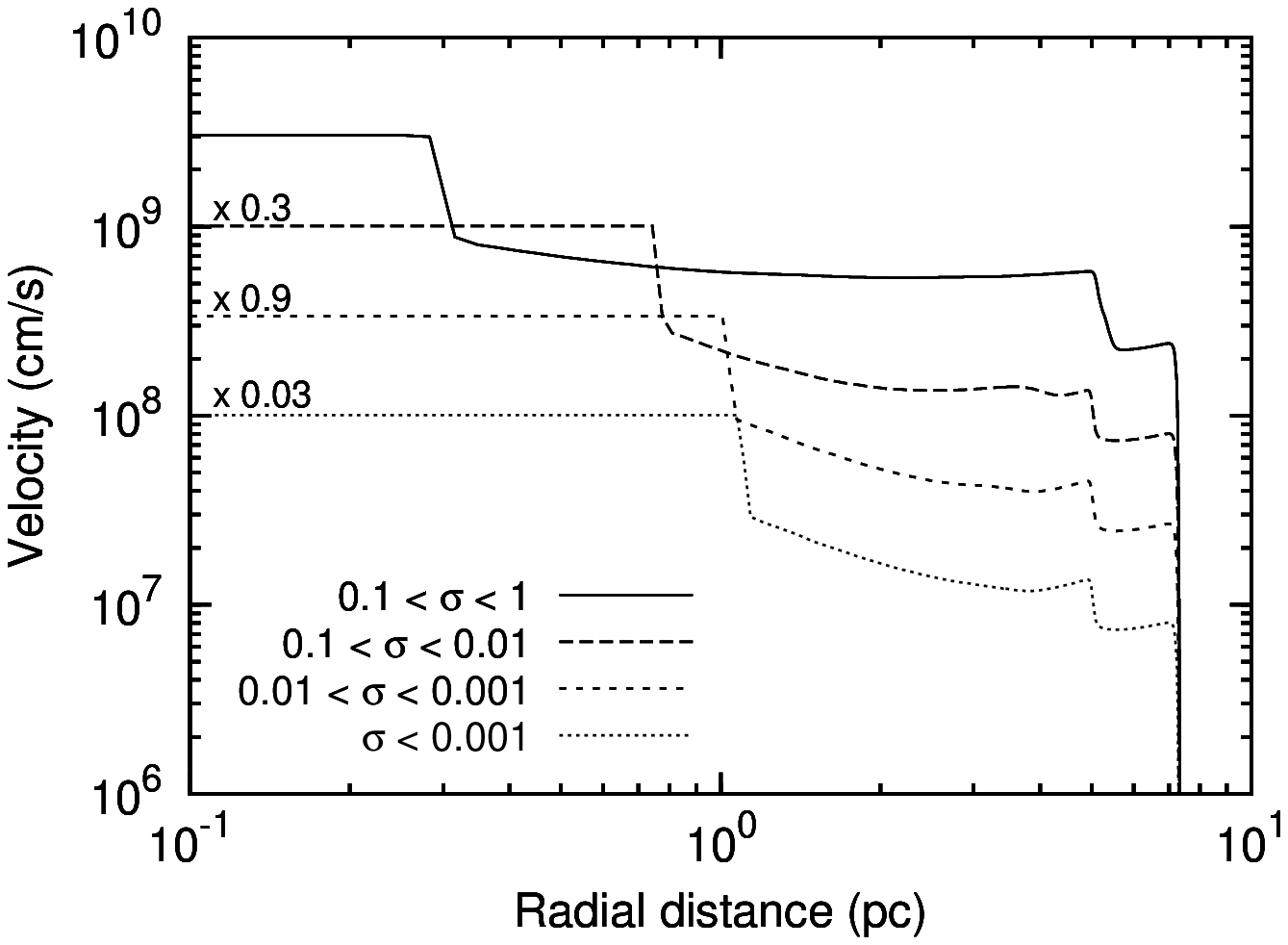}
   \caption{Radial profiles of the flow velocity corresponding to the scenarios in Fig \ref{bpres}.  To enhance clarity, the velocities for the various $\sigma$ values have been multiplied by different factors.  All velocities originally have a value of $c$ at the inner boundary of the computational domain.}
              \label{vpres}%
\end{minipage}
\end{figure*}

Figure \ref{bpres} shows the radial profile of the magnetic field for various values of $\sigma$.  Note that this value is not constant throughout the nebula, but varies with position.  The ranges given in the legend of the figure correspond to the possible values that $\sigma$ can assume for a given scenario, with $\sigma < 0.001$ representing the hydrodynamic/kinematic limit. Comparing the hydrodynamic limit with the $0.001\le \sigma\le 0.01$ scenario shows that increase in the magnetic pressure does not lead to a significant change in the size of the PWN.  Apart from a larger magnetic field, these two scenarios also predict a very similar magnetic field profile. A significant deviation from the hydrodynamic results only appears when $\sigma \ge 0.01$.   

When the magnetic pressure is relatively unimportant (i.e. a low $\sigma$ value), the compression ratio of the termination shock is large, and the magnetic field is enhanced across the shock.  Additionally, the flow speed decreases towards the boundary of the PWN, leading to an continual increase in $B$ as a function of position.  By contrast, a larger $\sigma$ value decreases the compression ratio, resulting in a post-shock magnetic field that decreases as a function of position, with the rate of decrease slightly smaller than the pre-shock value of $B\propto 1/r$. This decrease is caused by an acceleration of the post-shock fluid, in turn caused by the gradient of the magnetic pressure.

An increasing in magnetic pressure is sufficient to change the speed of the longitudinal pressure waves, leading to a change in the PWN evolution.  A stronger magnetic field at the inner boundary increase the size of the PWN, but also exerts a stronger pressure on $R_{\rm{ts}}$, pushing it closer to the pulsar. Note that the effect of magnetic pressure on the PWN evolution may vary when different assumptions are made regarding the initial conditions of the SNR and the boundary conditions of the PWN.

For large $\sigma$ values, the decreasing magnetic profile in Figure \ref{bpres} qualitatively agrees with the steady-state MHD solutions presented by \cite{Kennel_Coroniti}.  On the other hand, these authors found that the post-shock magnetic field increases to a maximum at a few termination shock radii, followed by a decrease towards the outer boundary of the PWN when $\sigma \sim 0.001$.  Panel 1 in Figure \ref{p4versameling2} shows that the kinematic approach predicts a similar character for magnetic field character in the inner part of the PWN, provided that the reverse shock has reached $R\s{pwn}$.  The main difference, compared to the steady-state results of \cite{Kennel_Coroniti}, is that the magnetic field in Figure \ref{p4versameling2} increases again in the outer parts of the PWN.  Spherically-symmetric MHD simulations, extending up to $t=1500$ yr, have also been performed by \cite{Bucciantini-etal-2004b} for the case $\sigma < 0.003$.  The results of these authors predict a magnetic field that increases with position, qualitatively similar to the results in Figure \ref{bpres}.   

For completeness, the corresponding radial profiles of the flow velocity are shown in Fig \ref{vpres}.  To improve visibility, the velocity of the various profiles have been artificially reduced, with the exception of the $0.001\le \sigma\le 0.01$ scenario.  The unaltered solutions all have a velocity of $c$ at the inner boundary.  From Figure \ref{vpres} it can be seen that the hydrodynamic limit and the $0.001\le \sigma\le 0.01$ have similar flow velocities and radial profiles.  Furthermore, when $0.1 \le \sigma \le 1$, the flow velocity is almost constant throughout the nebula, in agreement with the predictions made by \cite{Kennel_Coroniti}.

\section{Evolution in a Non-uniform Interstellar Medium}

\begin{figure*}[th]
   \centering
  \includegraphics[width=63pc]{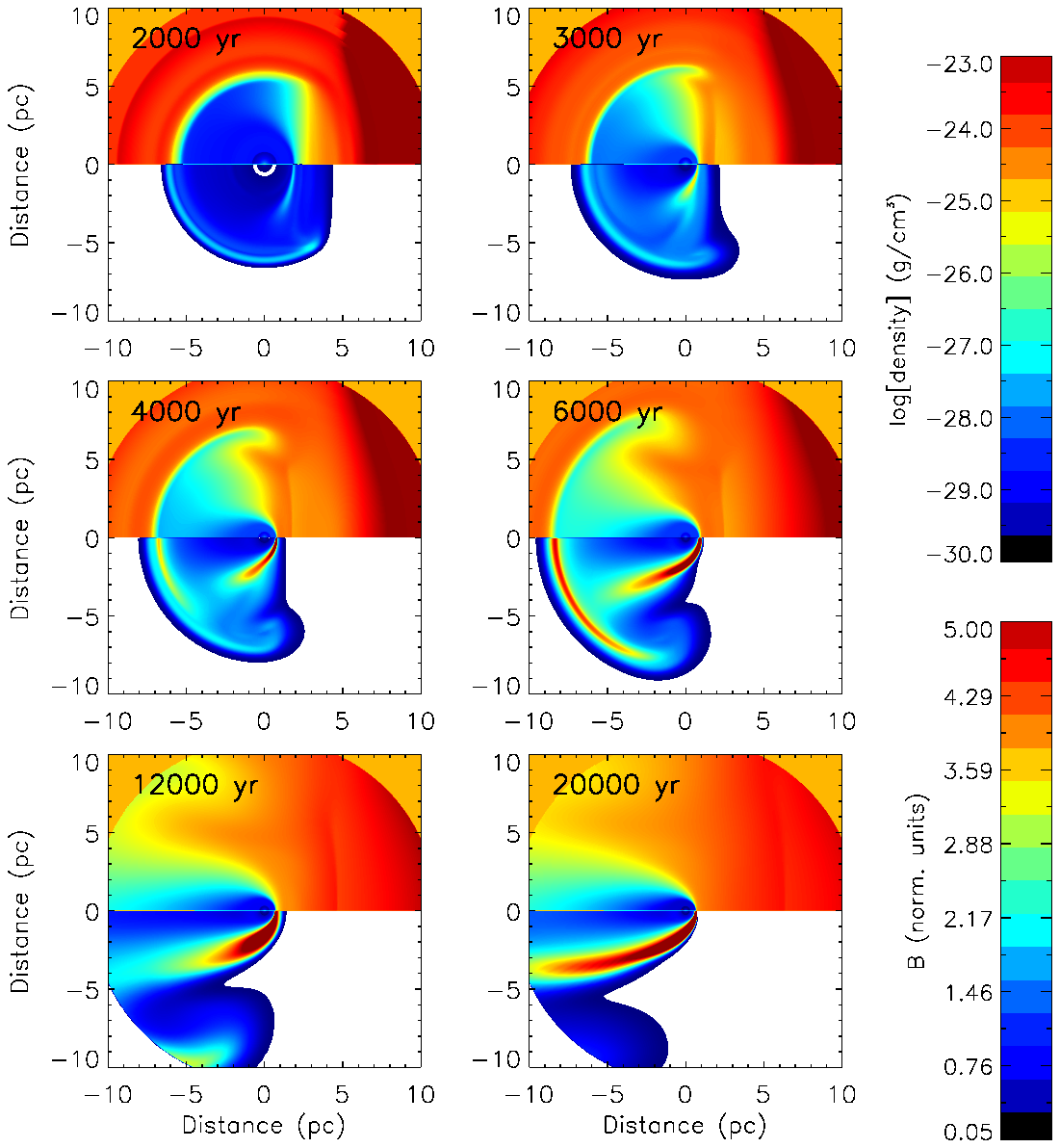}
   \caption{SNR-PWN system evolving in an interstellar medium with a density of $\rho_{\rm{ism}}=10^{-24}$ $\mbox{g cm}^{-3}$, with a higher density region,  $\rho_{\rm{ism}}=10^{-23}$ $\mbox{g cm}^{-3}$ located at $x=5$ pc.  The results correspond to the $M_{\rm{ej}}=3M_{\odot}$, $\tau$ = 3000 yr, and $L_{0}=10^{39}$ $\mbox{erg s}^{-1}$ scenario.  The top halves of the panels show the density, and the bottom halves the magnetic field.  The different panels represent snapshots of the evolution at various times.  The time shown in the top-left corner of the panels correspond to the time elapsed after the initial explosion.}
              \label{pa2}%
    \end{figure*}

\begin{figure*}[ht]
   \centering
  \includegraphics[width=63pc]{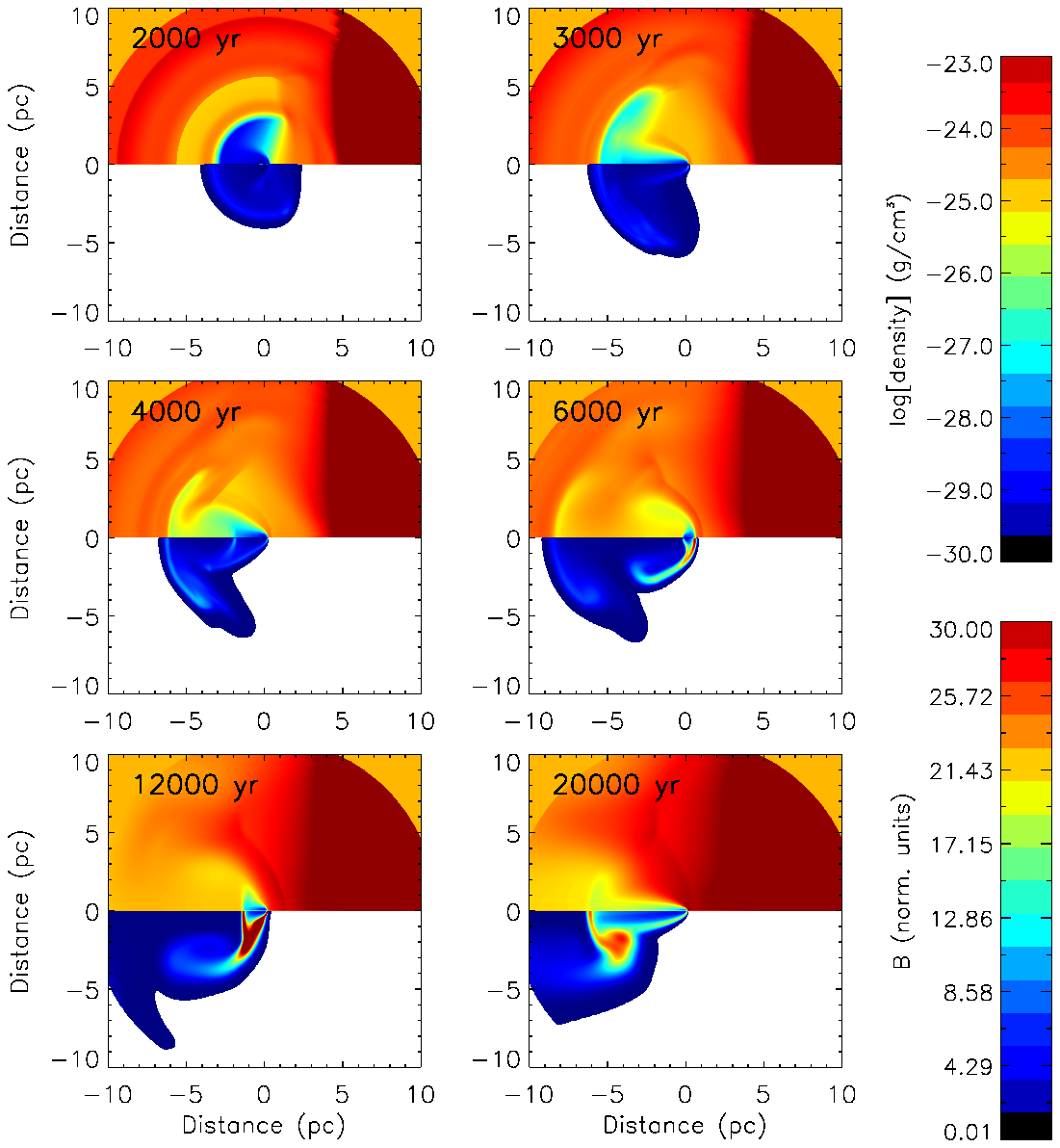}
  \vspace{-0.8cm}
   \caption{The same as Figure \ref{pa2}, except $L_{0}=10^{38}$ $\mbox{erg s}^{-1}$ and $\rho_{\rm{ism}}=10^{-21}$ $\mbox{g cm}^{-3}$ for the 'cloud'.}
              \label{pb3}%
    \end{figure*}

The ISM is very often inhomogeneous, and it is of particular interest to study the evolution of a composite remnant expanding into regions of varying density (e.g. due to the presence of a molecular cloud).  The aim is not to investigate any particular object or region, but rather to show some interesting end-result scenarios which can occur. The initial and boundary conditions for this scenario are identical to those described in Section 2, with the difference that the ISM has a discontinuous increase in density at a specified distance from the origin of the supernova explosion.  

It is well-known that an interaction between a shock front and a discontinuity (CD) separating fluids of different densities leads to a refraction of the shock \citep[e.g.][]{Colella1989}.  In the simplest case the original shock front will be refracted into a wave that travels into the transmission medium, and a wave that is reflected back into the incident medium \citep[e.g.][]{Colella1989}.  Depending on a number of factors, the transmitted and reflected waves can either be in the form of a shock, an expansion, or a number of complicated wave structures \citep[e.g.][]{Nourgaliev2005, Delmont2010}. One would therefore expect that the interaction of the forward shock of the SNR with the molecular cloud should lead to a similar wave pattern. 

Although difficult to predict the exact character of the refraction, \cite{Delmont2010} found that the most important parameters controlling the refraction are the angle of incidence between the shock front and CD, the Mach number of the initial shock in the incident medium, and the ratio of the densities on opposite sides of the CD.  Additionally, if a magnetic field is present, the value of the plasma $\beta$, can also influence the interaction \citep{Delmont2010}.

An important feature of the interaction of an oblique shock with a CD is the formation of Richtmeyr-Meshkov vorticity instabilities at the CD \citep[e.g.][]{Colella1989, Nourgaliev2005, Delmont2010}.  As previously discussed, the set-up used for the present numerical model is such that it does not allow for the development of Rayleigh-Taylor instabilities.  The same is also true for the development of fine-scale refraction patterns and Richtmeyr-Meshkov instabilities.  It is emphasised that the aim of the present numerical simulations is to obtain an understanding of the evolution of the large-scale structure of the PWN-SNR system in a non-uniform environment.     

For the scenario presented in Figure \ref{pa2}, the SNR evolves into an ISM with a density of $\rho_{\rm{ism}}=10^{-24}$ $\mbox{g cm}^{-3}$.  At $x=5$ pc from the position of the initial explosion, the density of the ISM is larger by a factor 10, i.e. $\rho_{\rm{ism}}=10^{-23}$ $\mbox{g cm}^{-3}$.  The mass of the ejecta is $M_{\rm{ej}}=3M_{\odot}$, the spin-down time-scale $\tau$ = 3000 yr, and the initial luminosity of the pulsar $L_{0}=10^{39}$ $\mbox{erg s}^{-1}$.   

After roughly $t=2000$ yr has elapsed, the forward shock reaches the denser region at $x = 5$ pc.  The sound speed in the incident ISM ($x < 5$ pc) is larger than the sound speed in the transmission ISM ($x \ge 5$ pc), and the interaction is thus classified as \emph{fast-slow} \citep[e.g.][]{Colella1989}.  A study of the radial profiles (not shown) shows that the interaction leads to a sub-sonic wave that is reflected back into the incident ISM, and a sub-sonic wave that is transmitted into the denser ISM.   When the reflection wave interacts with the PWN, the $t = 3000$ yr panel shows that the part of the PWN located at $x > 0$ pc is compressed and pushed back towards the pulsar (located at the origin).  As time evolves, the reverse shock flows over an increasingly larger part of the PWN, deforming initially spherical nebula into a bullet-shaped PWN.  

The initial compression, as well as the subsequent deformation, leads to a pile-up of the magnetic field in the nose of the PWN, as well as long the edge of the nebula.  After $t=4000$ yr, an additional enhancement of the magnetic field also becomes visible at the opposite edge of the PWN.  The PWN material transported away from the pulsar (in the $-x$ direction) by the reflection wave is decelerated by the supernova ejecta, leading to an additional pile-up of the magnetic field.  For times $t > 6000$ yr, the decrease in the pulsar's spin-down luminosity becomes important, leading to a pressure gradient between the PWN and ejecta/ISM.  A wind is created that continually flows over the PWN ($t=12000$ yr and $t=20000$ yr panels), resulting in a further compression of the magnetic field along the edge of the nebula.

For the second scenario presented in Figure \ref{pb3}, the ISM has a density $\rho_{\rm{ism}}=10^{-24}$ g/cm$^3$, with a factor thousand increase at $x=5$ pc, i.e. $\rho_{\rm{ism}}=10^{-21}$ g/cm$^3$.  The mass of the ejecta and spin-down time-scale is again chosen as $M_{\rm{ej}}=3M_{\odot}$ and $\tau$ = 3000 yr, respectively, while the initial luminosity is reduced by a factor ten to $L_{0}=10^{38}$ $\mbox{erg s}^{-1}$.  Note that the scale of the magnetic field in Figure \ref{pb3} is different to that of Figure \ref{pa2}.    

The interaction of the sub-sonic reflection wave with the PWN is significantly more pronounced, with the PWN being almost crushed by the wave.  A closer look at the $t=12 000$ yr panel reveals that some of the PWN material initially transported in the $-x$-direction (as a result of the reflection wave) is now flowing in the opposite direction.  The primary reflection wave flows through the lower density PWN, eventually reaching the discontinuity between the PWN and higher density supernova ejecta.  This results in the formation of a secondary reflection wave that propagates in the opposite direction.  This effect is not visible in Figure \ref{pa2} since a larger $L_0$ implies a larger pressure in the PWN, which would cancel out such an effect.  As the secondary wave propagates towards the pulsar, the magnetic field is dragged along with it, leading to a pile-up of the magnetic field.  The $t=20000$ yr panel shows that this magnetic cloud is eventually pushed in the $-x$-direction.  The largest magnetic field is therefore found in a region that is spatially uncorrelated with the position of the pulsar.

\section{Summary and Conclusions}

This paper investigates the evolution of a composite supernova remnant for a number of scenarios, with an emphasis on the evolution of the PWN.  A hydrodynamic model is used for the simulations, with the magnetic field included in a kinematic fashion.  In this approach the interaction between the fluid and magnetic field is strictly one-sided as the effect of the magnetic field on the flow is neglected.  The kinematic results were compared to a more correct treatment of the problem where the effect of the magnetic pressure is included.  It was found that the two approaches give similar results when the ratio of magnetic to particle energy is $\sigma < 0.01$.   

One of the aims of the paper is to determine the effect that the initial luminosity ($L_0$) and the spin-down time-scale ($\tau$) of the pulsar, together with the mass of the stellar ejecta ($M\s{ej}$) and density of the ISM ($\rho\s{ism}$) have on the evolution of a spherically-symmetric composite remnant expanding into a homogeneous ISM.  It was found that the evolution of the PWN is primarily determined by $L_0$ and $\rho\s{ism}$, while the influence of $M\s{ej}$ is almost negligible.  It was also found that $\tau$ only has an influence on the evolution when the spin-down time-scale is smaller than the time needed for the reverse shock to reach the outer boundary of the nebula, $R\s{pwn}$.

For the $L_0=10^{40}$ $\mbox{erg s}^{-1}$, $\tau=3000$ yr scenario, the interaction between the PWN and reverse shock only leads to a decrease in the expansion rate of $R\s{pwn}$.  For smaller $L_0$ values, the PWN undergoes a compression phase, with a decrease in $L_0$ leading to a larger compression.  Furthermore, a larger ISM density decreases the time needed by the reverse shock to reach $R\s{pwn}$, thereby markedly reducing the size of the PWN.

The parameter study further found that the termination shock radius, $R_{\rm{ts}}$, increases with time until the PWN encounters the reverse shock. The latter sends a reflection wave through the PWN towards the pulsar, resulting in a decrease of $R_{\rm{ts}}$.  The evolution of $R_{\rm{ts}}$ is an important quantity as it may help to determine the progress of the reverse shock through the PWN.  Additionally, the ratio of the PWN boundary to the termination shock radius, $R_{\rm{ts}} /R_{\rm{pwn}}$, has also previously been used  to derive the initial spin-period, $P_0$, for a number of pulsars \citep{Van-der-Swaluw-2003}.   

Studies of the evolution of the average magnetic field are important when attempting to understand so-called ''dark sources'', i.e. TeV $\gamma$-ray sources without a synchrotron counterpart \citep{H_Aharonian2008}. From Figure \ref{fig:b_field} it can be seen that a rapid expansion of the PWN ($L_0=10^{40}$ $\mbox{erg s}^{-1}$, $\tau=3000$ yr) leads to a significant decrease in the average magnetic field as a function of time, which in turn will lead to a faint synchrotron source.

The evolution of $R\s{pwn}$ and the average magnetic field are of particular importance to the one-zone (spatially-independent) radiation models that are often used to derive PWN parameters \citep[e.g.][]{Zhang2008, Fang2010, Tanaka2011}.  Additionally, the radial profiles of the magnetic field and flow velocity can also be incorporated into spatially-dependent transport models that describe the evolution of the non-thermal particle spectrum in the nebula (Vorster and Moraal 2013, accepted for publication).   

The simulations of a composite remnant evolving into an inhomogeneous ISM show that an initially spherical PWN will evolve into a bullet-shaped nebula.  This is in agreement with observations since most PWN are observed as offset from the pulsar \citep[for a summary of observations, see e.g.][]{Gaensler-2006,de-Jager-2009}.  The simulations further show that the interaction between the nebula and the asymmetric reverse shock (resulting from an inhomogeneous ISM) leads to an enhancement of the the magnetic field at the edge of the PWN.  Recent \emph{Suzaku} observations of the Vela PWN reveal an X-ray nebula that extends beyond the $3^{\circ} \times 2^{\circ}$ radio nebula \citep{Katsuda-2011}.  This is somewhat surprising as synchrotron radiation is expected to effectively cool the X-ray producing leptons, leading to an X-ray nebula that is smaller than the radio counterpart.  Although \citet{Katsuda-2011} favour a scenario in which these leptons effectively diffuse outward, the authors do not rule out a scenario in which lower energy leptons produce X-rays as a result of an increase in the magnetic field at the edge of the nebula.  

Decreasing the value of $L_0$, while simultaneously increasing the density gradient in the ISM, leads to the creation of a magnetic "cloud" in a region that is spatially separated from the pulsar.  The associated synchrotron emission will be enhanced, leading to a brighter region that is unconnected with the present position of the pulsar.

\begin{acknowledgements}
   The authors are grateful for partial financial support granted to them by the South African National Research Foundation (NRF) and by the Meraka Institute as part of the funding for the South African Centre for High Performance Computing (CHPC).
\end{acknowledgements}

\bibliographystyle{aa}
\bibliography{References}

\begin{thebibliography}{57}
\expandafter\ifx\csname natexlab\endcsname\relax\def\natexlab#1{#1}\fi

\bibitem[{{Aharonian} {et~al.}(2008){Aharonian}, {Akhperjanian}, {Barres de
  Almeida}, {Bazer-Bachi}, {Behera}, {Beilicke}, {Benbow}, {Bernl{\"o}hr},
  {Boisson}, {Bolz}, {Borrel}, {Braun}, {Brion}, {Brown}, {B{\"u}hler},
  {Bulik}, {B{\"u}sching}, {Boutelier}, {Carrigan}, {Chadwick}, {Chounet},
  {Clapson}, {Coignet}, {Cornils}, {Costamante}, {Dalton}, {Degrange},
  {Dickinson}, {Djannati-Ata{\"i}}, {Domainko}, {Drury}, {Dubois}, {Dubus},
  {Dyks}, {Egberts}, {Emmanoulopoulos}, {Espigat}, {Farnier}, {Feinstein},
  {Fiasson}, {F{\"o}rster}, {Fontaine}, {Funk}, {F{\"u}{\ss}ling}, {Gallant},
  {Giebels}, {Glicenstein}, {Gl{\"u}ck}, {Goret}, {Hadjichristidis}, {Hauser},
  {Hauser}, {Heinzelmann}, {Henri}, {Hermann}, {Hinton}, {Hoffmann}, {Hofmann},
  {Holleran}, {Hoppe}, {Horns}, {Jacholkowska}, {de Jager}, {Jung},
  {Katarzy{\'n}ski}, {Kendziorra}, {Kerschhaggl}, {Kh{\'e}lifi}, {Keogh},
  {Komin}, {Kosack}, {Lamanna}, {Latham}, {Lemi{\`e}re}, {Lemoine-Goumard},
  {Lenain}, {Lohse}, {Martin}, {Martineau-Huynh}, {Marcowith}, {Masterson},
  {Maurin}, {Maurin}, {McComb}, {Moderski}, {Moulin}, {de Naurois}, {Nedbal},
  {Nolan}, {Ohm}, {Olive}, {de O{\~n}a Wilhelmi}, {Orford}, {Osborne},
  {Ostrowski}, {Panter}, {Pedaletti}, {Pelletier}, {Petrucci}, {Pita},
  {P{\"u}hlhofer}, {Punch}, {Ranchon}, {Raubenheimer}, {Raue}, {Rayner},
  {Renaud}, {Ripken}, {Rob}, {Rolland}, {Rosier-Lees}, {Rowell}, {Rudak},
  {Ruppel}, {Sahakian}, {Santangelo}, {Schlickeiser}, {Sch{\"o}ck},
  {Schr{\"o}der}, {Schwanke}, {Schwarzburg}, {Schwemmer}, {Shalchi}, {Sol},
  {Spangler}, {Stawarz}, {Steenkamp}, {Stegmann}, {Superina}, {Tam},
  {Tavernet}, {Terrier}, {van Eldik}, {Vasileiadis}, {Venter}, {Vialle},
  {Vincent}, {Vivier}, {V{\"o}lk}, {Volpe}, {Wagner}, {Ward}, {Zdziarski}, \&
  {Zech}}]{H_Aharonian2008}
{Aharonian}, F., {Akhperjanian}, A.~G., {Barres de Almeida}, U., {et~al.} 2008,
  \aap, 477, 353

\bibitem[{{Aharonian} {et~al.}(2006){Aharonian}, {Akhperjanian}, {Bazer-Bachi},
  {Beilicke}, {Benbow}, {Berge}, {Bernl{\"o}hr}, {Boisson}, {Bolz}, {Borrel},
  {Braun}, {Breitling}, {Brown}, {Chadwick}, {Chounet}, {Cornils},
  {Costamante}, {Degrange}, {Dickinson}, {Djannati-Ata{\"i}}, {O'C.~Drury},
  {Dubus}, {Emmanoulopoulos}, {Espigat}, {Feinstein}, {Fontaine}, {Fuchs},
  {Funk}, {Gallant}, {Giebels}, {Glicenstein}, {Goret}, {Hadjichristidis},
  {Hauser}, {Hauser}, {Heinzelmann}, {Henri}, {Hermann}, {Hinton}, {Hofmann},
  {Holleran}, {Horns}, {Jacholkowska}, {de Jager}, {Kh{\'e}lifi}, {Klages},
  {Komin}, {Konopelko}, {Latham}, {Le Gallou}, {Lemi{\`e}re},
  {Lemoine-Goumard}, {Lohse}, {Martin}, {Martineau-Huynh}, {Marcowith},
  {Masterson}, {McComb}, {de Naurois}, {Nedbal}, {Nolan}, {Noutsos}, {Orford},
  {Osborne}, {Ouchrif}, {Panter}, {Pelletier}, {Pita}, {P{\"u}hlhofer},
  {Punch}, {Raubenheimer}, {Raue}, {Rayner}, {Reimer}, {Reimer}, {Ripken},
  {Rob}, {Rolland}, {Rowell}, {Sahakian}, {Saug{\'e}}, {Schlenker},
  {Schlickeiser}, {Schuster}, {Schwanke}, {Siewert}, {Sol}, {Spangler},
  {Steenkamp}, {Stegmann}, {Superina}, {Tavernet}, {Terrier}, {Th{\'e}oret},
  {Tluczykont}, {van Eldik}, {Vasileiadis}, {Venter}, {Vincent}, {V{\"o}lk}, \&
  {Wagner}}]{H_Aharonian2006}
{Aharonian}, F., {Akhperjanian}, A.~G., {Bazer-Bachi}, A.~R., {et~al.} 2006,
  \aap, 449, 223

\bibitem[{{Arzoumanian} {et~al.}(2002){Arzoumanian}, {Chernoff}, \&
  {Cordes}}]{Arzoumanian2002}
{Arzoumanian}, Z., {Chernoff}, D.~F., \& {Cordes}, J.~M. 2002, \apj, 568, 289

\bibitem[{{Axford}(1972)}]{Axford-1972}
{Axford}, W.~I. 1972, in Solar Wind, ed. C.~P. {Sonett}, P.~J. {Coleman}, \&
  J.~M. {Wilcox}, 598

\bibitem[{{Blondin} {et~al.}(2001){Blondin}, {Chevalier}, \&
  {Frierson}}]{Blondin-etal-2001}
{Blondin}, J.~M., {Chevalier}, R.~A., \& {Frierson}, D.~M. 2001, ApJ, 563, 806

\bibitem[{{Blondin} \& {Ellison}(2001)}]{Blondin-and-Ellison-2001}
{Blondin}, J.~M. \& {Ellison}, D.~C. 2001, ApJ, 560, 244

\bibitem[{{Blondin} {et~al.}(1998){Blondin}, {Wright}, {Borkowski}, \&
  {Reynolds}}]{Blondin-etal-1998}
{Blondin}, J.~M., {Wright}, E.~B., {Borkowski}, K.~J., \& {Reynolds}, S.~P.
  1998, \apj, 500, 342

\bibitem[{{Borkowski} {et~al.}(1997){Borkowski}, {Blondin}, \&
  {McCray}}]{Borkowski-etal-1997}
{Borkowski}, K.~J., {Blondin}, J.~M., \& {McCray}, R. 1997, ApJ, 477, 281

\bibitem[{{Bucciantini}(2002)}]{Bucciantini-2002}
{Bucciantini}, N. 2002, A\&A, 387, 1066

\bibitem[{{Bucciantini} {et~al.}(2004{\natexlab{a}}){Bucciantini}, {Amato},
  {Bandiera}, {Blondin}, \& {Del Zanna}}]{Bucciantini-etal-2004a}
{Bucciantini}, N., {Amato}, E., {Bandiera}, R., {Blondin}, J.~M., \& {Del
  Zanna}, L. 2004{\natexlab{a}}, A\&A, 423, 253

\bibitem[{{Bucciantini} {et~al.}(2005){Bucciantini}, {Amato}, \& {Del
  Zanna}}]{Bucciantini-etal-2005}
{Bucciantini}, N., {Amato}, E., \& {Del Zanna}, L. 2005, A\&A, 434, 189

\bibitem[{{Bucciantini} {et~al.}(2004{\natexlab{b}}){Bucciantini}, {Bandiera},
  {Blondin}, {Amato}, \& {Del Zanna}}]{Bucciantini-etal-2004b}
{Bucciantini}, N., {Bandiera}, R., {Blondin}, J.~M., {Amato}, E., \& {Del
  Zanna}, L. 2004{\natexlab{b}}, A\&A, 422, 609

\bibitem[{{Bucciantini} {et~al.}(2003){Bucciantini}, {Blondin}, {Del Zanna}, \&
  {Amato}}]{Bucciantini-etal-2003}
{Bucciantini}, N., {Blondin}, J.~M., {Del Zanna}, L., \& {Amato}, E. 2003,
  A\&A, 405, 617

\bibitem[{{Chevalier}(1977)}]{Chevalier-1977}
{Chevalier}, R.~A. 1977, \araa, 15, 175

\bibitem[{{Chevalier}(2005)}]{Chevalier-2005}
{Chevalier}, R.~A. 2005, \apj, 619, 839

\bibitem[{{Colella} {et~al.}(1989){Colella}, {Henderson}, \&
  {Puckett}}]{Colella1989}
{Colella}, P., {Henderson}, L.~F., \& {Puckett}, E.~G. 1989, in 9th AIAA
  Computational Fluid Dynamics Conference, ed. D.~B. {Bliss} \& W.~O. {Miller},
  426--439

\bibitem[{Cranfill(1971)}]{Cranfill-1971}
Cranfill, C. 1971, PhD thesis, University of California, San Diego

\bibitem[{{De Jager} \& {Djannati-Ata{\"i}}(2009)}]{de-Jager-2009}
{De Jager}, O.~C. \& {Djannati-Ata{\"i}}, A. 2009, in Astrophysics and Space
  Science Library, ed. W.~{Becker}, Vol. 357, 451

\bibitem[{{Decourchelle} {et~al.}(2000){Decourchelle}, {Ellison}, \&
  {Ballet}}]{Decourchelle-etal-2000}
{Decourchelle}, A., {Ellison}, D.~C., \& {Ballet}, J. 2000, ApJ, 543, L57

\bibitem[{{Del Zanna} {et~al.}(2004){Del Zanna}, {Amato}, \&
  {Bucciantini}}]{Del-Zanna-etal-2004}
{Del Zanna}, L., {Amato}, E., \& {Bucciantini}, N. 2004, A\&A, 421, 1063

\bibitem[{{Delmont}(2010)}]{Delmont2010}
{Delmont}, P. 2010, PhD thesis, Arenberg Doctoral School of Sciene, Engineering
  and Technology, Katholieke Unversiteit, Leuven

\bibitem[{{Dwarkadas} \& {Chevalier}(1998)}]{Dwarkadas-and-Chevalier-1998}
{Dwarkadas}, V.~V. \& {Chevalier}, R.~A. 1998, ApJ, 497, 807

\bibitem[{{Ellison} {et~al.}(2004){Ellison}, {Decourchelle}, \&
  {Ballet}}]{Ellison-etal-2004}
{Ellison}, D.~C., {Decourchelle}, A., \& {Ballet}, J. 2004, A\&A, 413, 189

\bibitem[{{Fang} \& {Zhang}(2010)}]{Fang2010}
{Fang}, J. \& {Zhang}, L. 2010, \aap, 515, A20

\bibitem[{{Faucher-Gigu{\`e}re} \& {Kaspi}(2006)}]{Kaspi_birth}
{Faucher-Gigu{\`e}re}, C.-A. \& {Kaspi}, V.~M. 2006, ApJ, 643, 332

\bibitem[{{Ferreira} \& {De Jager}(2008)}]{Ferreira-and-de-Jager-2008}
{Ferreira}, S.~E.~S. \& {De Jager}, O.~C. 2008, A\&A, 478, 17

\bibitem[{{Ferreira} \& {Scherer}(2006)}]{Ferreira-and-Scherer-2006}
{Ferreira}, S.~E.~S. \& {Scherer}, K. 2006, ApJ, 642, 1256

\bibitem[{{Gaensler} \& {Slane}(2006)}]{Gaensler-2006}
{Gaensler}, B.~M. \& {Slane}, P.~O. 2006, \araa, 44, 17

\bibitem[{{Horns} {et~al.}(2006){Horns}, {Aharonian}, {Santangelo}, {Hoffmann},
  \& {Masterson}}]{Horns-2006}
{Horns}, D., {Aharonian}, F., {Santangelo}, A., {Hoffmann}, A.~I.~D., \&
  {Masterson}, C. 2006, \aap, 451, L51

\bibitem[{{Jun} \& {Jones}(1999)}]{Jun-and-Jones-1999}
{Jun}, B.-I. \& {Jones}, T.~W. 1999, ApJ, 511, 774

\bibitem[{{Kargaltsev} \& {Pavlov}(2008)}]{Kargaltsev-2008}
{Kargaltsev}, O. \& {Pavlov}, G.~G. 2008, in AIP Conference Series, Vol. 983,
  40 Years of Pulsars: Millisecond Pulsars, Magnetars and More, ed. C.~{Bassa},
  Z.~{Wang}, A.~{Cumming}, \& V.~M. {Kaspi}, 171--185

\bibitem[{{Katsuda} {et~al.}(2011){Katsuda}, {Mori}, {Petre}, {Yamaguchi},
  {Tsunemi}, {Bocchino}, {Bamba}, {Miceli}, {Hewitt}, {Temim}, {Uchida}, \&
  {Yoshii}}]{Katsuda-2011}
{Katsuda}, S., {Mori}, K., {Petre}, R., {et~al.} 2011, PASJ, 63, 827

\bibitem[{{Kennel} \& {Coroniti}(1984)}]{Kennel_Coroniti}
{Kennel}, C.~F. \& {Coroniti}, F.~V. 1984, ApJ, 283, 694

\bibitem[{{LeVeque}(2002)}]{Leveque-2002}
{LeVeque}, R.~J. 2002, {Finite Volume Methods for Hyperbolic Problems}
  ({Cambridge University Press})

\bibitem[{{McKee}(1974)}]{McKee1974}
{McKee}, C.~F. 1974, \apj, 188, 335

\bibitem[{{McKee} \& {Truelove}(1995)}]{McKee-and-Truelove-1995}
{McKee}, C.~F. \& {Truelove}, J.~K. 1995, Phys.Rep., 256, 157

\bibitem[{{Nourgaliev} {et~al.}(2005){Nourgaliev}, {Sushchikh}, {Dinh}, \&
  {Theofanous}}]{Nourgaliev2005}
{Nourgaliev}, R.~R., {Sushchikh}, S.~Y., {Dinh}, T.~N., \& {Theofanous}, T.~G.
  2005, Intern. J. Multiphase Flow, 31, 969

\bibitem[{{Rees} \& {Gunn}(1974)}]{Rees-and-Gunn-1974}
{Rees}, M.~J. \& {Gunn}, J.~E. 1974, \mnras, 167, 1

\bibitem[{{Reynolds} \& {Chevalier}(1984)}]{Reynolds-and-Chevalier-1984}
{Reynolds}, S.~P. \& {Chevalier}, R.~A. 1984, ApJ, 278, 630

\bibitem[{{Scherer} \& {Ferreira}(2005)}]{Scherer-and-Ferreira-2005a}
{Scherer}, K. \& {Ferreira}, S.~E.~S. 2005, Astrophys. \& Space Sci. Trans., 1,
  17

\bibitem[{{Sedov}(1959)}]{Sedov-1959}
{Sedov}, L.~I. 1959, {Similarity and Dimensional Methods in Mechanics}
  (Academic Press)

\bibitem[{{Shigeyama} \& {Nomoto}(1990)}]{Shigeyama-1990}
{Shigeyama}, T. \& {Nomoto}, K. 1990, \apj, 360, 242

\bibitem[{{Tanaka} \& {Takahara}(2011)}]{Tanaka2011}
{Tanaka}, S.~J. \& {Takahara}, F. 2011, \apj, 741, 40

\bibitem[{{Tang} \& {Wang}(2005)}]{Tang-and-Wang-2005}
{Tang}, S. \& {Wang}, Q.~D. 2005, ApJ, 628, 205

\bibitem[{{Tenorio-Tagle} {et~al.}(1991){Tenorio-Tagle}, {Rozyczka}, {Franco},
  \& {Bodenheimer}}]{Tenorio-Tagle-etal-1991}
{Tenorio-Tagle}, G., {Rozyczka}, M., {Franco}, J., \& {Bodenheimer}, P. 1991,
  MNRAS, 251, 318

\bibitem[{{Thornton} {et~al.}(1998){Thornton}, {Gaudlitz}, {Janka}, \&
  {Steinmetz}}]{Thornton-etal-1998}
{Thornton}, K., {Gaudlitz}, M., {Janka}, H.-T., \& {Steinmetz}, M. 1998, \apj,
  500, 95

\bibitem[{{Trac} \& {Pen}(2003)}]{Trac-and-Pen-2003}
{Trac}, H. \& {Pen}, U.-L. 2003, PASP, 115, 303

\bibitem[{{Truelove} \& {McKee}(1999)}]{Truelove-and-McKee-1999}
{Truelove}, J.~K. \& {McKee}, C.~F. 1999, ApJ, 120, 299

\bibitem[{{Van der Swaluw}(2003b)}]{Van-der-Swaluw-2003}
{Van der Swaluw}, E. 2003b, A\&A, 404, 939

\bibitem[{{Van der Swaluw} {et~al.}(2003){Van der Swaluw}, {Achterberg},
  {Gallant}, {Downes}, \& {Keppens}}]{Van-der-Swaluw-etal-2003}
{Van der Swaluw}, E., {Achterberg}, A., {Gallant}, Y.~A., {Downes}, T.~P., \&
  {Keppens}, R. 2003, A\&A, 397, 913

\bibitem[{{Van der Swaluw} {et~al.}(2001){Van der Swaluw}, {Achterberg},
  {Gallant}, \& {T{\'o}th}}]{Van-der-Swaluw-etal-2001}
{Van der Swaluw}, E., {Achterberg}, A., {Gallant}, Y.~A., \& {T{\'o}th}, G.
  2001, A\&A, 380, 309

\bibitem[{{Van der Swaluw} {et~al.}(2004){Van der Swaluw}, {Downes}, \&
  {Keegan}}]{Van-der-Swaluw-etal-2004}
{Van der Swaluw}, E., {Downes}, T.~P., \& {Keegan}, R. 2004, A\&A, 420, 937

\bibitem[{{Van der Swaluw} \& {Wu}(2001)}]{Van-der-Swaluw-and-Wu-2001}
{Van der Swaluw}, E. \& {Wu}, Y. 2001, ApJ, 555, L49

\bibitem[{{Woltjer}(1972)}]{Woltjer-1972}
{Woltjer}, L. 1972, ARA\&A, 10, 129

\bibitem[{{Woosley} \& {Janka}(2005)}]{Woosley-2005}
{Woosley}, S. \& {Janka}, T. 2005, Nature Physics, 1, 147

\bibitem[{{Zank}(1999)}]{Zank-1999}
{Zank}, G.~P. 1999, Space Sci. Rev., 89, 413

\bibitem[{{Zhang} {et~al.}(2008){Zhang}, {Chen}, \& {Fang}}]{Zhang2008}
{Zhang}, L., {Chen}, S.~B., \& {Fang}, J. 2008, \apj, 676, 1210

\end{thebibliography}

\end{document}